\documentclass[aps,prl,reprint,superscriptaddress,longbibliography,nobalancelastpage,groupedaddress]{revtex4-2}
\usepackage{graphicx}
\usepackage{dcolumn}
\usepackage{amsmath,amssymb}
\usepackage{latexsym}
\usepackage{mathrsfs}
\usepackage[usenames,dvipsnames]{color}
\usepackage{hyperref}
\usepackage{xcolor}

\usepackage[normalem]{ulem}
\newcommand{\be}{\begin{equation}} 
\newcommand{\ee}{\end{equation}} 
\newcommand{\bea}{\begin{eqnarray}} 
\newcommand{\eea}{\end{eqnarray}} 

\newcommand{\figref}[2]{[Fig.~\hyperref[#1]{\ref*{#1}(#2)}]}
\newcommand{\figrefi}[2]{[Fig.~\hyperref[#1]{\ref*{#1}(#2)}, inset]}
\newcommand{\textfigref}[2]{Fig.~\hyperref[#1]{\ref*{#1}(#2)}}
\newcommand{\textfigureref}[2]{Figure~\hyperref[#1]{\ref*{#1}(#2)}}

\newcommand{\textwholefigref}[1]{Fig.~\ref{#1}}

\newcommand{\figrefp}[2]{\hyperref[#1]{\ref*{#1}(#2)}}

\definecolor{linkcolor}{HTML}{223096}
\hypersetup{colorlinks,allcolors=linkcolor}

\bibpunct{\textcolor{linkcolor}{[}}{\textcolor{linkcolor}{]}}{\textcolor{linkcolor}{,}}{n}{}{;}
\renewcommand{\eqref}[1]{\hyperref[#1]{(\ref*{#1})}}

\begin{document}
\title{Elasticity and plasticity of epithelial gap closure}

\author{Maryam Setoudeh}
\affiliation{Max Planck Institute for the Physics of Complex Systems, N\"othnitzer Stra\ss e 38, 01187 Dresden, Germany}
\affiliation{\smash{Max Planck Institute of Molecular Cell Biology and Genetics, Pfotenhauerstra\ss e 108, 01307 Dresden, Germany}}
\affiliation{Center for Systems Biology Dresden, Pfotenhauerstra\ss e 108, 01307 Dresden, Germany}
\author{Pierre A. Haas}
\email{haas@pks.mpg.de}
\affiliation{Max Planck Institute for the Physics of Complex Systems, N\"othnitzer Stra\ss e 38, 01187 Dresden, Germany}
\affiliation{\smash{Max Planck Institute of Molecular Cell Biology and Genetics, Pfotenhauerstra\ss e 108, 01307 Dresden, Germany}}
\affiliation{Center for Systems Biology Dresden, Pfotenhauerstra\ss e 108, 01307 Dresden, Germany}
\date{\today}
\begin{abstract}
Epiboly, during which a tissue closes around the surface of the egg, pervades animal development. This epithelial gap closure involves cell intercalations at the edge of the gap. Here, inspired by serosa closure in the beetle \emph{Tribolium}, we study the interplay between these plastic cell rearrangements and the elasticity of the tissue in a minimal continuum model of the closure of a circular gap bounded by a contractile actomyosin cable. We discover two different closure mechanisms at the tissue scale depending on the energy barrier $E_\text{b}$ to and the energy $\Delta E$ released by intercalation: If $E_\text{b}\gg\Delta E$, cells intercalate into the gap to close it. For a fluidised tissue in which $E_\text{b}\ll\Delta E$, however, cells deintercalate from the boundary into the bulk of the tissue, and we reveal an emergent mechanical role of inhomogeneities of the actomyosin cable. Our work thus explains the mechanical role of tissue fluidisation in \emph{Tribolium} serosa closure and processes of epiboly and wound healing more generally.
\end{abstract}\maketitle
\begin{figure}[hb!]
\centering
\includegraphics[width=7.7cm]{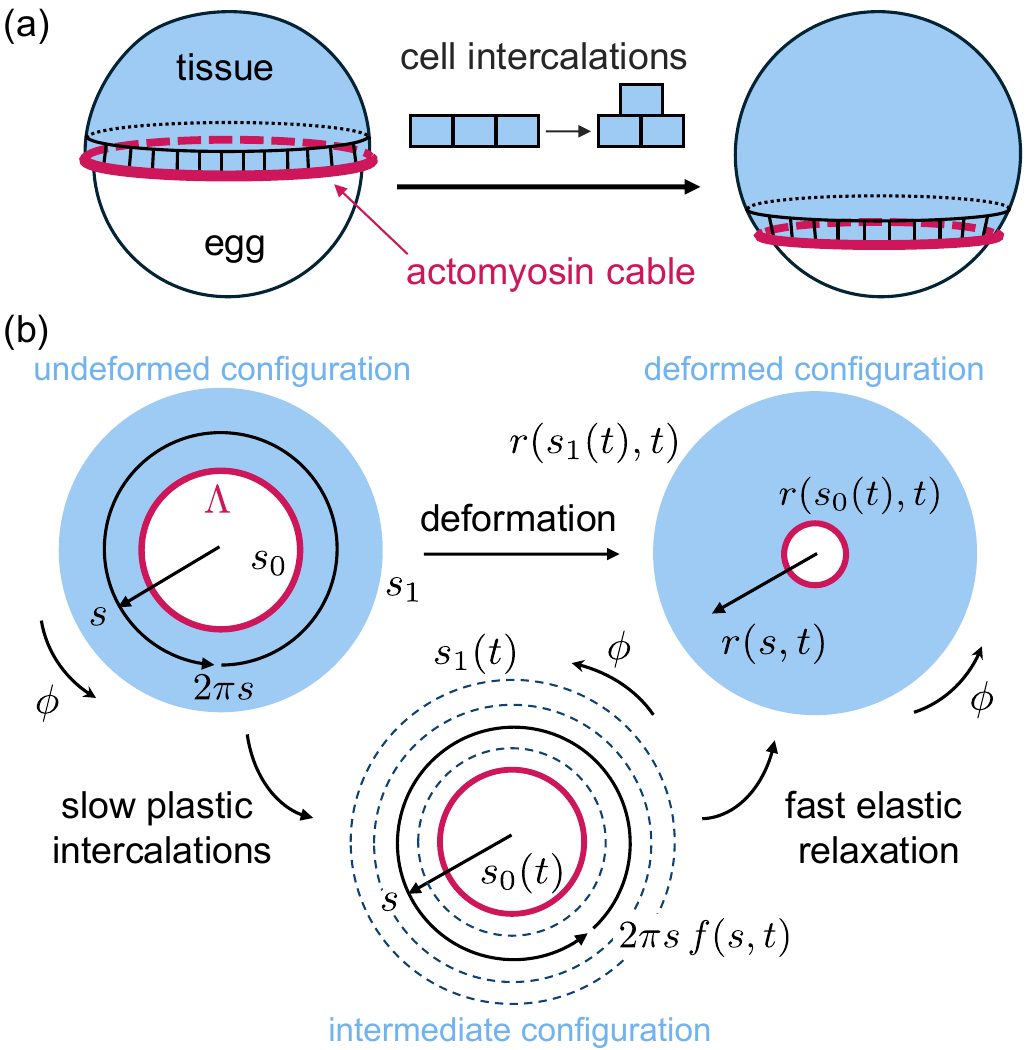}
\caption{Mechanics of epiboly. (a) Schematic of epiboly: a tissue closes around the surface of an egg. Epiboly is associated with a contractile actomyosin cable at the leading edge of tissue. During epiboly, the number of cells at the tissue boundary changes due to cell intercalations. (b) Minimal model of epiboly: a circular gap of radius $s_0$ closes in a concentric circular flat tissue of radius $s_1$ (top left). The gap is surrounded by a cable with uniform contractility $\Lambda$. The axisymmetric deformation decomposes into slow plastic intercalations and fast elastic relaxation. The intermediate configuration after intercalation (bottom) is described by the intercalation stretch $f(s,t)$ and the radii $s_0(t)$, $s_1(t)$ of the gap and the tissue. Points at radius $s$ in the undeformed configuration are at radius $r(s,t)$ in the deformed configuration (top right). See text for further explanation.}
\label{fig1}
\end{figure}

\renewcommand{\floatpagefraction}{.999}
One of the most conserved morphogenetic tissue movements during development is epiboly~\cite{solnica05}, the spreading and closing of an epithelium over the egg~\figref{fig1}{a}. The best-studied instantiation of this epithelial gap closure is probably epiboly in fish, during which tissue spreading is mediated by a contractile actomyosin cable at the leading edge of the tissue and cell division, cell intercalations, and cell shape changes within the tissue~\cite{keller87,koppen06,lepage10,behrndt12,campinho13,bruce20}. In particular, cell intercalations at the leading edge reduce the number of cells there to accommodate the change of the length of the leading edge during epiboly~\figref{fig1}{a}.

A particularly simple version of epiboly arises in the beetle \emph{Tribolium} \cite{benton13,jain20}, in which an extraembryonic tissue, the serosa, spreads over the embryo. Unlike zebrafish epiboly, there are no cell divisions in the serosa~\cite{jain20}. During serosa closure, the tissue fluidises near its leading edge, where an inhomogeneous actomyosin cable contracts the tissue~\cite{jain20}.

The mechanical basis for epiboly and the role of tissue fluidisation in this important morphogenetic process remain unclear, however, as does, specifically, the mechanical role of the inhomogeneity of the actomyosin cable in \emph{Tribolium} epiboly. This is because continuum descriptions of the interplay of tissue geometry and elasticity and the cell intercalations within the tissue are lacking. 

Borrowing the language of foams~\cite{cantat}, these cell intercalations, during which the network of cell-cell contacts changes, are T1 transitions. Mechanically, they are plastic rearrangements associated with crossing an energy barrier and the release of elastic energy after crossing said barrier~\cite{marmottant09,bi14,krajnc18,popovic21,haas25}. The vanishing of this energy barrier defines tissue fluidisation. The large amount of recent interest, in the physics \mbox{literature~\cite{krajnc18,popovic21,bi15,yan19,krajnc20,krajnc21,duclut22,yamamoto22,jain24,staddon25}}, in the mechanisms and phase transitions of tissue fluidisation and, more biologically, the importance of tissue fluidisation, beyond epiboly, in development and disease~\cite{hannezo22,lenne22} stress the need for coarse-grained descriptions of plastic cell intercalations in tissue mechanics.

Here, we describe the dynamics of this interplay of elasticity and plasticity in a minimal geometry of epithelial gap closure~\figref{fig1}{b}: a circular gap surrounded by a contractile cable closes axisymmetrically in a circular, flat tissue. By coarse-graining a discrete model of intercalations within the tissue, we discover how the energy barrier to and the elastic energy released by intercalations control the gap closure dynamics at the tissue scale.

Initially, the gap has radius $s_0$, and the tissue has radius $s_1$~\figref{fig1}{b}. Points at radius $s$ in this undeformed configuration end up at radius $r(s,t)$ in the deformed configuration. As in the general continuum theories of nonlinear plasticity~\cite{lubliner,bonet} and morphoelasticity~\cite{goriely}, this deformation decomposes multiplicatively into plastic intercalations and elastic relaxation~\figref{fig1}{b}. Intercalations reduce or increase the number of cells in each ring surrounding the gap and hence change the rest length of the ring at radius $s$ from $2\pi s$ in the undeformed configuration to $2\pi s\,f(s,t)$ in the intermediate configuration \figref{fig1}{b}, where $f(s,t)$ is an intercalation stretch. Due to cell intercalations across the tissue boundaries, the intrinsic radii of the gap and of the tissue are $s_0(t)$ and $s_1(t)$ in the intermediate configuration, respectively~\figref{fig1}{b}.

The timescale of plastic intercalations is much slower than the timescale of elastic relaxation; in \emph{Tribolium} epiboly for example, these are of the order of minutes and seconds, respectively~\cite{jain20}. Hence, at each instant in time, the elastic part of the deformation is determined by minimising the mechanical energy 
\begin{align}
\mathcal{E}(t)=2\pi\int_{s_0(t)}^{s_1(t)}{e(s,t)\,s\,\mathrm{d}s}+2\pi\Lambda r(s_0(t),t).\label{eq:E}
\end{align}
The first term is the elastic energy; the second term is the contribution from the cable, with uniform contractility $\Lambda$. The elastic energy density in Eq.~\eqref{eq:E} is
\begin{align}
e(s,t)=\dfrac{C}{2}\bigl[E_s(s,t)^2\!+\!E_s(s,t)E_\phi(s,t)\!+\!E_\phi(s,t)^2\bigr]f(s,t)\label{eq:ed}
\end{align}
for axisymmetric deformations. We derive this expression in the Supplemental Material~\footnote{See Supplemental Material at [url to be inserted], which includes Refs.~\cite{goriely,steigmann13,haas21,ramachandran24,lubliner,bonet,erbay97,dervaux08,ogden,dervaux09,abramowitz,nr}, for (i)~the derivation of the elastic energy density, (ii) details of the calculation of the intercalation rates, and (iii)~details of the numerical solution of the governing equations, including additional simulation results.}. Here, $C$ is an elastic modulus and the elastic strains are\nocite{steigmann13,haas21,ramachandran24,erbay97,dervaux08,ogden,dervaux09,abramowitz,nr}
\begin{align}
&E_s(s,t)=\dfrac{\partial r}{\partial s}(s,t)-1,&&E_\phi(s,t)=\dfrac{r(s,t)}{s\,f(s,t)}-1.\label{eq:strains}
\end{align}

\begin{figure}[t]
\centering\includegraphics[width=\linewidth]{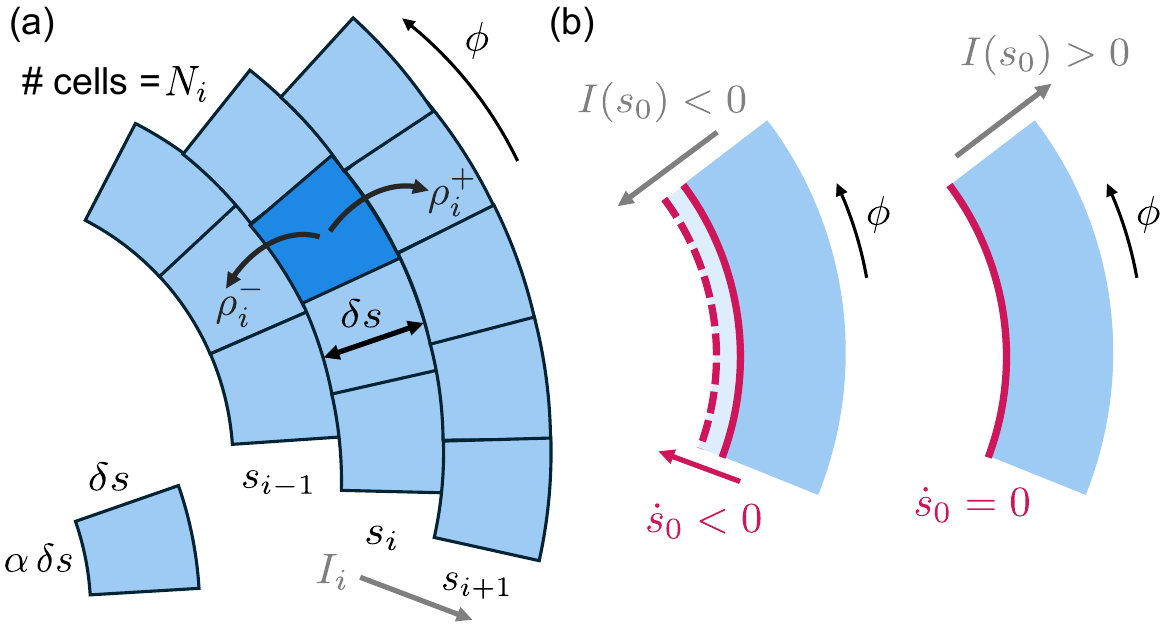}
\caption{Intercalation dynamics. (a) Discrete model of cell intercalations in the bulk of the tissue: each of the $N_i$ cells in a ring of radius $s_i$ can intercalate outwards or inwards, with rates $\rho^\pm_i$, to the rings of radii ${s_{i\pm 1}=s_i\pm\delta s}$. This defines the effective intercalation rate $I_i$ from $s_i$ to $s_{i+1}$. Inset: each cell has aspect ratio~$\alpha=2\pi\beta$. (b) Intercalation at the (inner) boundary: if $I(s_0)<0$, then $\dot{s}_0<0$; if $I(s_0)>0$, then $\dot{s}_0=0$.}\label{fig2}
\end{figure}

To close the model, we need to impose the slow dynamics of the intercalation stretch $f(s,t)$ and of the boundaries $s_0(t)$, $s_1(t)$. For this purpose, we coarse-grain a minimal discrete model of intercalations~\figref{fig2}{a} in which the intermediate configuration of the sheet consists of concentric rings of cells of width $\delta s$. The ring of radius~$s_i$ thus has intrinsic length $2\pi s_if_i(t)$, where ${f_i(t)=f(s_i,t)}$. The discrete cells have undeformed aspect ratio~$\alpha$ \figrefi{fig2}{a}, so that the number of cells in this ring is ${N_i(t)=[2\pi s_if_i(t)]/(\alpha\,\delta s)}$. Each cell in this ring can intercalate outwards or inwards, to the rings of radii $s_{i\pm 1}=s_i\pm\delta s$, with respective rates $\rho_i^\pm(t)$ \figref{fig2}{a}. The effective rate at which cells intercalate from $s_i$ to $s_{i+1}$ is therefore ${I_i(t)\propto N_i(t)\rho_i^+(t)-N_{i+1}(t)\rho_{i+1}^-(t)}$. With these definitions, ${N_i(t+\delta t)-N_i(t)=[I_{i-1}(t)-I_i(t)]\,\delta t}$, so, taking the continuum limit $\delta t\to 0,\delta s\to 0$,
\begin{subequations}
\begin{align}
\dfrac{\partial f}{\partial t}=-\dfrac{\beta}{s}\dfrac{\partial I}{\partial s},
\end{align}
where $\beta=\alpha/2\pi$. This holds in the bulk of the tissue, for $s_0(t)<s<s_1(t)$. At the tissue boundaries, it extends to 
\begin{align}
\dfrac{\partial f}{\partial t}=-\dfrac{\beta}{s}\dfrac{\partial I}{\partial s}+\Delta_0(t)\delta\bigl(s-s_0(t)\bigr)+\Delta_1(t)\delta\bigl(s-s_1(t)\bigr),\label{eq:df}
\end{align}
\end{subequations}
in which the amplitudes $\Delta_0(t),\Delta_1(t)$ of the Dirac deltas are determined by cell number conservation, as are the dynamics $\dot{s}_0(t)$, $\dot{s}_1(t)$: In the absence of cell divisions, as in \emph{Tribolium} epiboly~\cite{jain20}, for any ${S>s_0(t)}$, the rate of change of the number of cells in $s_0(t)<s<S$ balances intercalation across $s=S$. Thus
\begin{subequations}
\begin{align}
\dfrac{\mathrm{d}}{\mathrm{d}t}\left(\int_{s_0(t)}^{S}{f(s,t)\,s\,\mathrm{d}s}\right)+\beta I(S)=0.
\end{align}
Using Eq.~\eqref{eq:df}, this reduces to
\begin{align}
\beta I(s_0(t),t)+\dfrac{s_0(t)}{2}\Delta_0(t)=\dot{s}_0(t)s_0(t)f(s_0(t),t).\label{eq:cnc}
\end{align}
\end{subequations}
As illustrated in \textfigref{fig2}{b}, if $I(s_0(t),t)<0$, then cells intercalate into the gap at the inner boundary, so ${\dot{s}_0(t)<0}$. Thus ${\dot{s}_0(t)=\beta I(s_0(t),t)/[s_0(t)f(s_0(t),t)]<0}$, ${\Delta_0(t)\!=\!0}$ satisfy Eq.~\eqref{eq:cnc}. By contrast, if ${I(s_0(t),t)>0}$, then cells deintercalate from the boundary into the bulk of the tissue. This reduces the number of cells in the innermost ring of cells, but does not remove it. Hence its radius in the intermediate configuration remains unchanged, i.e., $\dot{s}_0(t)=0$, although elastic relaxation can of course still change the deformed radius $r(s_0(t),t)$. Equation~\eqref{eq:cnc} then implies ${\Delta_0(t)=-2\beta I(s_0(t),t)/s_0(t)}$. An analogous argument at the outer boundary ${s=s_1(t)}$ shows that ${\dot{s}_1(t)=\beta I(s_1(t),t)/[s_1(t)f(s_1(t),t)]>0}$ and $\Delta_1(t)=0$ if $I(s_1(t),t)>0$, but if $I(s_1(t),t)<0$, then $\dot{s}_1(t)=0$ and $\Delta_1(t)=-2\beta I(s_1(t),t)/s_1(t)$. In this way, the intercalation dynamics break time-reversal symmetry at the boundaries of the gap.

\begin{figure}[t!]
\centering
\includegraphics[width=\linewidth]{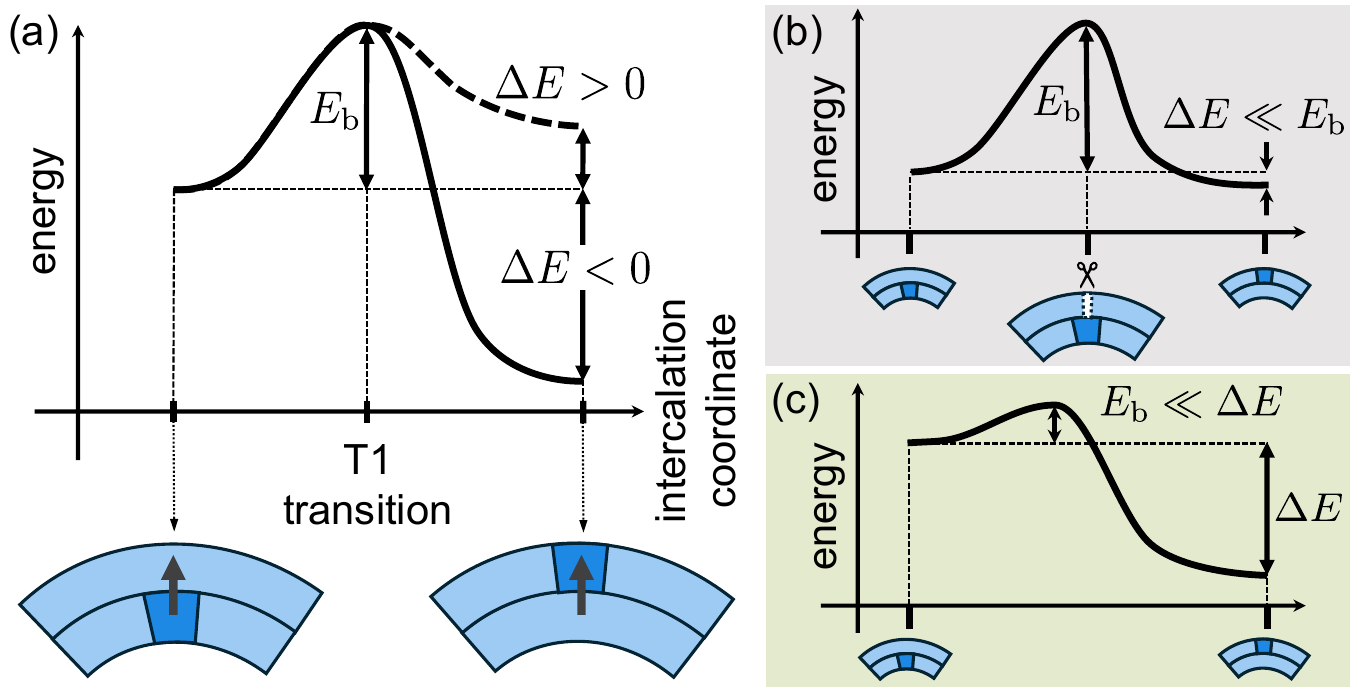}
\caption{Intercalation energetics. (a) Sketch of the energy landscape near a T1 transition (insets), illustrating the energy barrier $E_\text{b}>0$ at the T1 transition point and the elastic energy released ($\Delta E<0$) or gained ($\Delta E>0$) during intercalation. (b) Energy landscape for $\Delta E\ll E_\text{b}$ (high energy barrier). Insets: model of a high energy barrier, resulting from opening a gap in the tissue during intercalation. (c) Energy landscape for $E_\text{b}\ll\Delta E$ (fluidised tissue).}\label{fig3}
\end{figure}

We are left to determine the dependence of the intercalation rates $I_i(t)$ on the mechanics, for which we need to impose the single-cell intercalation rates $\rho_i^\pm(t)$. Intercalation involves crossing a barrier $E_\text{b}>0$ in the energy landscape at the T1 transition point~\figref{fig3}{a}. During the T1 transition, the system releases ($\Delta E<0$) or gains ($\Delta E>0$) elastic energy~\figref{fig3}{a}. We therefore impose
\begin{align}
\rho_i^\pm\propto\exp{\left(-\dfrac{E_{\text{b},i}^\pm+\Delta E_i^\pm}{\theta}\right)},\label{eq:rho}
\end{align}
in which the effective temperature $\theta$ represents, e.g., active tension fluctuations~\cite{krajnc18,duclut22,krajnc20,krajnc21,yamamoto22,duclut21} driving intercalations. A similar model, inspired by the Eyring model of chemical kinetics~\cite{espenson}, has been deployed previously on T1 transitions in tissues~\cite{marmottant09}. 

Here, $N_i\to N_i-1$ and $N_{i\pm 1}\to N_{i\pm 1}+1$ in an intercalation from $s_i$ to $s_{i\pm 1}=s_i\pm\delta s$, so ${f_i\to f_i-\beta\,\delta s/s_i}$ and ${f_{i\pm 1}\to f_{i\pm 1}+\beta\,\delta s/s_{i\pm 1}}$, with, on Taylor expanding, ${f_{i\pm 1}=f_i\pm \delta s\,f'(s_i)+\tfrac{1}{2}\delta s^2f''(s_i)+O\bigl(\delta s^3\bigr)}$. During this intercalation, the elastic part of the deformation, i.e., $r(s,t)$, is fixed, so we consider the elastic energy density in Eq.~\eqref{eq:ed} as $e(s,f(s))$. Hence

\begin{widetext}
\vspace{-14pt}
\begin{align}
\Delta E_i^\pm&=2\pi\,\delta s\left\{\left[e\left(s_i,f_i-\dfrac{\beta\,\delta s}{s_i}\right)s_i+e\left(s_{i\pm 1},f_{i\pm 1}+\dfrac{\beta\,\delta s}{s_{i\pm 1}}\right)s_{i\pm 1}\right]-\bigl[e(s_i,f_i)s_i+e(s_{i\pm1},f_{i\pm 1})s_{i\pm 1}\bigr]\right\}\nonumber\\
&=\delta s^3\left[\dfrac{\beta^2}{s}\dfrac{\partial^2e}{\partial f^2}\pm\beta\dfrac{\mathrm{d}}{\mathrm{d}s}\left(\dfrac{\partial e}{\partial f}\right)\right]_{\substack{s=s_i\\f=f_i}}+O\bigl(\delta s^4\bigr),\quad\text{where }\dfrac{\partial e}{\partial f}=\dfrac{C}{2}\Bigl(E_s^2-E_\phi^2-E_s-2E_\phi\Bigr),\;\dfrac{\partial^2 e}{\partial f^2}=\dfrac{Cr^2}{s^2f^3}>0,\label{eq:DeltaE}
\end{align}\vspace{-8pt}
\end{widetext}
using Eqs.~\eqref{eq:ed} and \eqref{eq:strains}, as detailed in the Supplemental Material~\cite{Note1}. Unlike this energy released or gained during intercalation, the energy barrier $E_\text{b}$ is not determined by the tissue-scale mechanics in the elastic model: it is in general a property of the mechanics at the cell scale. In what follows, we therefore focus on two limiting cases.

We begin with the case $E_\text{b}\gg\Delta E$ of a high energy barrier \figref{fig3}{b}. We impose a minimal model of the energy barrier~[\textfigref{fig3}{b}, insets]: a constant cost of opening up cell junctions during intercalation sets $E_\text{b}$. To ensure $E_\text{b}\gg\Delta E$, we choose $E^\pm_{\smash{\text{b},i}}=B\,\delta s^2$. Importantly, from Eq.~\eqref{eq:rho}, this implies $\rho_i^+=\rho_i^-$ at leading order. Hence subleading terms contribute to the leading-order expression for $I_i\propto N_i\rho_i^+-N_{i+1}\rho_{i+1}^-$. Physically, this means that even though $E_\text{b}\gg\Delta E$, $\Delta E$ cannot be neglected, and hence time-reversal symmetry remains broken, also in the bulk of the tissue. 

To take the continuum limit, we still need to impose the asymptotic scaling of the effective temperature, $\theta=\vartheta\,\delta s^\tau$. In the ``cold'' case (large~$\tau$), there are no intercalations for $E_\text{b}$ cannot be overcome. In the ``hot'' case (small $\tau$), intercalations are driven independently of the mechanics. Here, we therefore analyse the interesting intermediate case, which is $\tau=2$, as shown in the Supplemental Material~\cite{Note1}. With this, we obtain~\cite{Note1}
\begin{align}
I\propto -\left[\dfrac{\mathrm{d}}{\mathrm{d}s}(sf)+\dfrac{2\beta}{\vartheta}sf\dfrac{\mathrm{d}}{\mathrm{d}s}\left(\dfrac{\partial e}{\partial f}\right)\right]\mathrm{e}^{-B/\vartheta}.\label{eq:Ihigh}
\end{align}
This finally couples the intercalation rate to the mechanics, via the expression for $\partial e/\partial f$ given by Eq.~\eqref{eq:DeltaE}, and hence closes the model. The energy barrier merely rescales the intercalation rate (or, equivalently, time) and the aspect ratio $\alpha=2\pi\beta$ only rescales $\vartheta$.

Next, we turn to the case $E_\text{b}\ll\Delta E$ of a fluidised tissue~\figref{fig3}{c}, in which the intermediate-temperature case corresponds to $\tau=3$ and we find~\cite{Note1}
\begin{align}
I\propto -s f\exp{\left(-\dfrac{\beta^2}{s\vartheta}\dfrac{\partial^2 e}{\partial f^2}\right)\sinh{\left(\dfrac{\beta}{\vartheta}\dfrac{\mathrm{d}}{\mathrm{d}s}\left(\dfrac{\partial e}{\partial f}\right)\right)}},\label{eq:Ifluid}
\end{align}
in which $\partial e/\partial f$, $\partial^2e/\partial f^2$ are given in Eq.~\eqref{eq:DeltaE}. Strikingly, $\beta$ does not merely rescale $\vartheta$ in this case: there is an emergent role of the microscopic cell geometry at the macroscopic scale of the fluidised tissue.

\begin{figure*}[t!]
\centering\includegraphics[width=\linewidth]{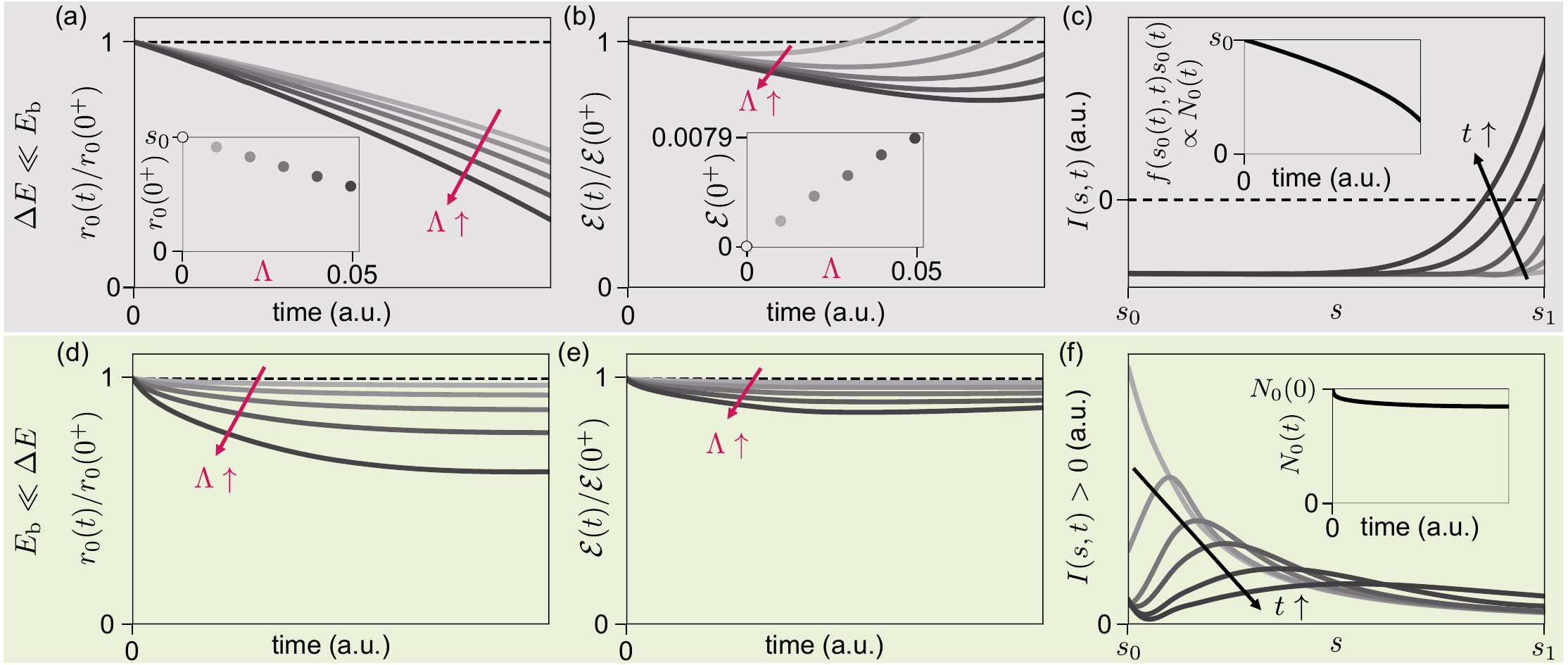}
\caption{Dynamics of epithelial gap closure. Top row: dynamics for a high energy barrier ($\Delta E\ll E_\text{b}$). (a) Plot of the inner radius $r_0(t)\equiv r(s_0(t),t)$, scaled with its initial contracted value $r_0(0^+)$ before intercalations, against time $t$, for increasing values of contractility $\Lambda$. Inset: plot of $r_0(0^+)$ against $\Lambda$. (b) Plot of energy $\mathcal{E}(t)$, scaled with $\mathcal{E}(0^+)$, against $t$, for increasing values of $\Lambda$. Inset: plot of $\mathcal{E}(0^+)$ against $\Lambda$. (c) Plot of the intercalation rate $I(s,t)$ against $s$, at increasing times $t$. Inset: decrease of $f(s_0(t),t)s_0(t)$, proportional to the number $N_0(t)$ of cells surrounding the gap, with time. Bottom row: analogous plots for the fluidised model ($\Delta E\gg E_{\text{b}}$). (d) Plot of $r_0(t)$. (e) Plot of $\mathcal{E}(t)$. (f) Plot of $I(s,t)$ against~$s$. Inset: plot of $N_0(t)$. Parameter values: $s_0(0)=0.2$, $s_1(0)=1$, $C=\vartheta=1$, $B=0.1$, $\beta=0.02$, [panels (c), (f)] $\Lambda=0.02$. a.u.: arbitrary units.}\label{fig4}
\end{figure*}

We analyse the intercalation dynamics in these two cases numerically, by integrating Eq.~\eqref{eq:df} in \textsc{Matlab} (The MathWorks, Inc.) and solving the boundary value problem associated with Eq.~\eqref{eq:E} at each timestep using its \texttt{bvp4c} solver, as explained in detail in the Supplemental Material~\cite{Note1}. Here, we discuss the case of a fixed tissue radius, $r(s_1(t),t)=s_1(0)$, but note that a force-free boundary condition at $s=s_1(t)$ gives similar results~\cite{Note1}.

We begin by observing that the radius of the gap decreases in the case of a high energy barrier [${\Delta E\ll E_\text{b}}$, \textfigref{fig4}{a}]. At higher contractility $\Lambda$, not only is the contracted radius before intercalations smaller \figrefi{fig4}{a}, but also the gap shrinks faster \figref{fig4}{a}. At early times, passive intercalations dissipate energy \figref{fig4}{b}. Unsurprisingly, this dissipation is faster at higher $\Lambda$, because the elastic energy before intercalations increases with~$\Lambda$ \figrefi{fig4}{b}. At later times, however, energy starts to increase~\figref{fig4}{b}. This stresses that energetically unfavourable active intercalations at the boundaries of the tissue can persist to build up stresses in the tissue. Importantly, the intercalation rate is negative near the gap~\figref{fig4}{c}, so these dynamics are driven by cells intercalating into the gap. As a result, the number of cells surrounding the gap decreases~\figrefi{fig4}{c}.

\begin{figure}
\centering
\includegraphics[width=\linewidth]{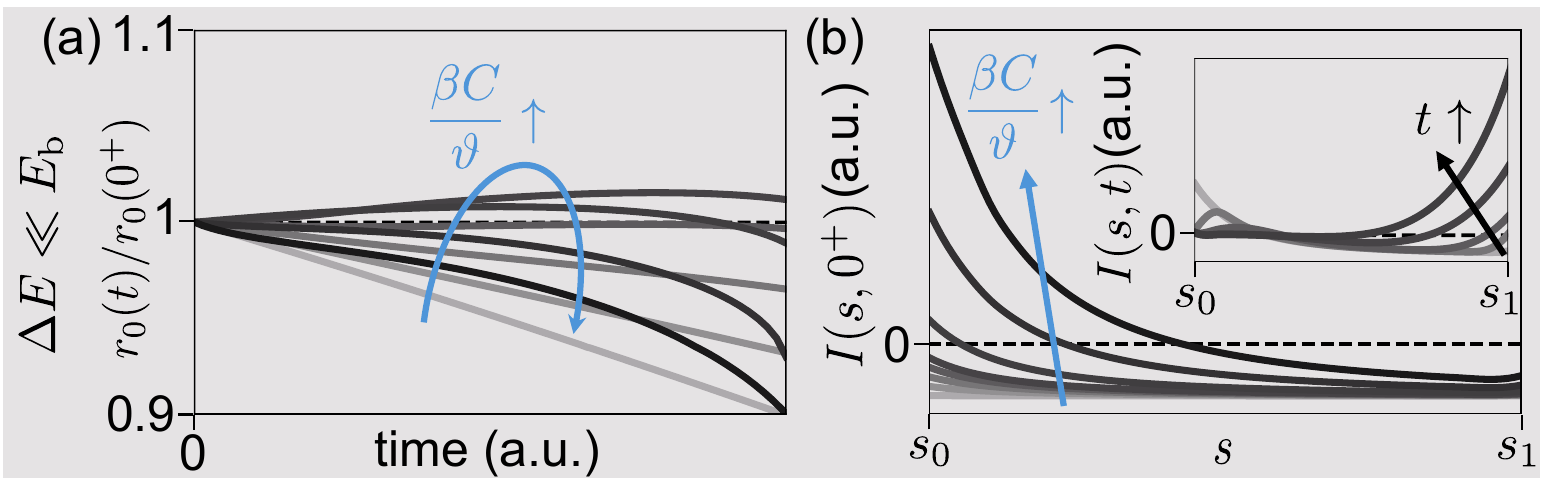}
\caption{Dynamics of epiboly for $\Delta E\ll E_\text{b}$. Effect of $\beta C/\vartheta$ on (a) $r_0(t)$ and (b) the initial intercalation rate $I(s,0^+)$. Inset of panel (b): Plot of $I(s,t)$ against $s$ for increasing times $t$ at large $\beta C/\vartheta$. Other parameter values as for \textwholefigref{fig4}.}\label{fig5}
\end{figure}

For a fluidised tissue [$\Delta E\gg E_\text{b}$, \textfigref{fig4}{d}], the radius of the gap decreases, too, but appears to converge later. Unlike the previous case, the energy starts to increase earlier at higher $\Lambda$~\figref{fig4}{e}. Strikingly however, the intercalation dynamics at the cell scale are different: the intercalation rate is positive near the gap~\figref{fig4}{f}, so cells deintercalate from the gap into the bulk of the tissue, whence the number of cells surrounding the gap decreases~\figrefi{fig4}{f}.

Equation~\eqref{eq:Ihigh} suggests that outward intercalation at the inner boundary is also possible for $E_\text{b}\gg\Delta E$ if $\beta C/\vartheta$ is large. This leads, however, to transient widening of the gap \figref{fig5}{a} as the intercalation rate at the inner boundary increases~\figref{fig5}{b} and, also, to more intercalation at the outer boundary than at the inner boundary at late times~\figrefi{fig5}{b}.

Thus only the fluidised model is compatible with the observed dynamics of \emph{Tribolium} epiboly, where cells deintercalate from the boundary into the bulk~\cite{jain20}. This is consistent with the fluidisation of the \emph{Tribolium} serosa near its boundary inferred previously from cell shapes~\cite{jain20}.

The inhomogeneity of the contractile cable surrounding the \emph{Tribolium} serosa leads to inhomogeneous cell aspect ratios $\alpha=2\pi\beta$. To describe their averaged effect, we replace, in Eq.~\eqref{eq:Ifluid}, ${\beta\to\langle\beta\rangle}$, ${\beta^2\to\langle\beta^2\rangle}$, where angle brackets denote averages. By convexity~\cite{stirzaker}, ${\langle\beta^2\rangle<\langle \beta\rangle^2}$ for an inhomogeneous cable. Since $\partial^2e/\partial f^2>0$ from Eq.~\eqref{eq:DeltaE}, this means that $|I|$ is increased compared to the case of a homogeneous cable. In other words, the inhomogeneous cable surrounding the \emph{Tribolium} serosa favours its closure. This effect requires fluidisation of the serosa, because the intercalation rate for a nonfluidised tissue, based on Eq.~\eqref{eq:Ihigh}, depends only on~$\langle \beta\rangle$, but not on $\langle\beta^2\rangle$.

In summary, we have shown how the interplay of elasticity and plasticity drives epithelial gap closure. We have discovered how the microscopic energetics of intercalation at the cell scale can give rise to very different macroscopic dynamics at the tissue scale. Our work explains the role of tissue fluidisation and inhomogeneities of the contractile cable in \emph{Tribolium} serosa closure. More generally, it provides a continuum mechanical framework for understanding epiboly and related processes of wound healing~\cite{brugues14,vedula15,begnaud16,babu24,lim24,movrin25}, which additionally involve active cell migration, but for which the role of tissue fluidisation has been recognised, too~\cite{tetley19,sarate24}.

Our minimal model predicts that a gap in a fluidised tissue shrinks, but does not close fully at constant contractility~\figref{fig4}{d}. How strengthening of contractile cables~\cite{vedula15} or stresses exerted by the tissues surrounding the \emph{Tribolium} serosa contribute to overcoming this limit is an open biological question. Meanwhile, extending our continuum framework to three-dimensional intercalations, described by a tensorial intercalation stretch, is an important challenge for future theoretical work. 

\begin{acknowledgments}
\emph{Acknowledgments}---We thank Sifan Yin and Carl Modes for comments on the manuscript, and Akanksha Jain and Pavel Toman\v{c}\'ak for discussions about the biology of \emph{Tribolium} epiboly. We gratefully acknowledge funding from the Max Planck Society.

\emph{Data availability}---\textsc{Matlab} code for the solution of the governing equations is openly available at Ref.~\footnote{Code is available at \href{https://doi.org/10.5281/zenodo.17062741}{\texttt{zenodo:17062742}}.}.
\end{acknowledgments}

\bibliography{references}
\end{document}


\renewcommand{\theequation}{S\arabic{equation}}
\renewcommand{\thefigure}{S\arabic{figure}}
\renewcommand{\thepage}{S\arabic{page}}
\renewcommand{\thetable}{S}
\title{Elasticity and plasticity of epithelial gap closure\\\textbullet{} Supplemental Material \textbullet{}}

\author{Maryam Setoudeh}
\affiliation{Max Planck Institute for the Physics of Complex Systems, N\"othnitzer Stra\ss e 38, 01187 Dresden, Germany}
\affiliation{\smash{Max Planck Institute of Molecular Cell Biology and Genetics, Pfotenhauerstra\ss e 108, 01307 Dresden, Germany}}
\affiliation{Center for Systems Biology Dresden, Pfotenhauerstra\ss e 108, 01307 Dresden, Germany}
\author{Pierre A. Haas}
\affiliation{Max Planck Institute for the Physics of Complex Systems, N\"othnitzer Stra\ss e 38, 01187 Dresden, Germany}
\affiliation{\smash{Max Planck Institute of Molecular Cell Biology and Genetics, Pfotenhauerstra\ss e 108, 01307 Dresden, Germany}}
\affiliation{Center for Systems Biology Dresden, Pfotenhauerstra\ss e 108, 01307 Dresden, Germany}\maketitle

This Supplemental Material divides into three parts. In the first part, we derive the elastic energy density for axisymmetric deformations of an elastic sheet. In the second part, we give the details of the calculation of the intercalation rates and discuss the role of the effective temperature. In the final part, we provide details of the numerical solution of the governing equations.

\section{Derivation of the elastic energy density}
In this section, we derive the energy of an incompressible neo-Hookean axisymmetric elastic sheet [Eq.~(2) of the main text]. The sheet remains flat throughout its deformation. In its undeformed configuration, it has thickness $\epsilon h$, where $\epsilon \ll 1$ is an asymptotic parameter in which we will expand the full three-dimensional elasticity and plasticity of the sheet. Our derivation is based on similar calculations for morphoelastic shells~\cite{steigmann13,haas21} and rods~\cite{ramachandran24}, with our notation based on that of Refs.~\cite{haas21,ramachandran24}.
\subsection{Geometry of an axisymmetric elastic sheet}
\begin{figure}[b]
    \centering
\includegraphics[width=\linewidth]{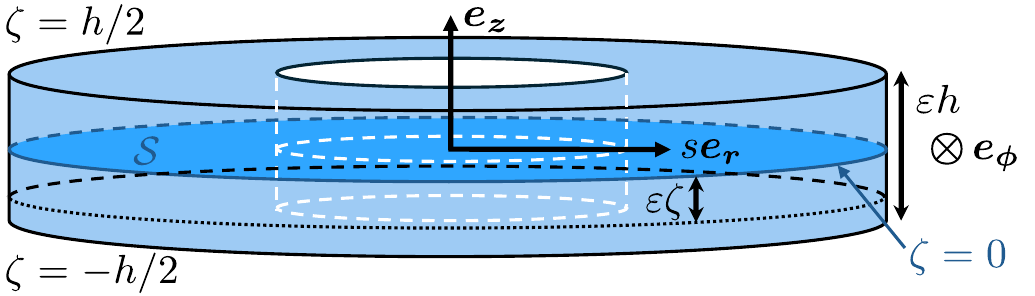}
    \caption{Undeformed configuration $\mathcal{V}$ of an axisymmetric sheet of thickness $\varepsilon h$ with respect to the basis $\vec{e_r},\vec{e_\phi},\vec{e_z}$ of cylindrical polars. The midsurface is $\mathcal{S}$, on which $\zeta=0$, where $\zeta\in[-h/2,h/2]$ is the transverse coordinate.}
    \label{figS1}
\end{figure}
We begin by deriving the geometry of the different configurations of the sheet. We start with the undeformed configuration $\mathcal{V}$ of the sheet~\wholefigref{figS1}, which we describe in terms of its midsurface $\mathcal{S}$. With respect to the basis $\mathcal{B}=\{\vec{e_{r}},\vec{e_{\phi}},\vec{e_{z}}\}$ of cylindrical coordinates, a point on $\mathcal{S}$ has position vector
\begin{align}
    \vec{\rho}(s,\phi) = s \vec{e_{r}}(\phi),
\end{align}
where $s$ is the radial coordinate and $\phi$ is the azimuthal coordinate. The position of a point in $\mathcal{V}$ is therefore
\begin{align}
    \vec{r}(s,\phi , \zeta) = \vec{\rho}(s,\phi) + \epsilon \zeta \vec{e_z}.\label{eq:tbundef}
\end{align}
where $\zeta\in[-h/2,h/2]$ is the transverse coordinate~\wholefigref{figS1}. The tangent basis of $\mathcal{V}$ is therefore 
\begin{align}
&\dfrac{\partial \vec{r}}{\partial s} = \vec{e}_{r},
&&\dfrac{\partial \vec{r}}{\partial \phi} = s \vec{e}_{\phi},
&&\dfrac{\partial \vec{r}}{\partial \zeta} = \epsilon \vec{e}_{z},\label{eq:drundef}
\end{align}
with corresponding scale factors
\begin{align}
    \chi_s = 1 , && \chi_{\phi} = s, && \chi_\zeta = \epsilon.
\end{align}
In particular, the volume element of $\mathcal{V}$ is
\begin{align}
\mathrm{d}V =\chi_{s}\chi_{\phi}\chi_{\zeta}\,\mathrm{d}s\,\mathrm{d}\phi\,\mathrm{d}\zeta = \epsilon s\,\mathrm{d}s\,\mathrm{d}\phi\,\mathrm{d}\zeta. \label{eq:dvundef}
\end{align}
As the sheet deforms, $\mathcal{V}$ maps to its deformed configuration~$\tilde{\mathcal{V}}$. Under this mapping, $\mathcal{S}$ maps to the deformed midsurface $\tilde{\mathcal{S}}$, with position vector
 \begin{align}
     \vec{\tilde{\rho}}(s,\phi) = \tilde{s}(s)\vec{e}_{r}(\phi),
 \end{align}
where $\tilde{s}(s)$ is the deformed radial coordinate~\footnote{The deformed radial coordinate is $r(s)$ in the main text and in the remaining sections of this Supplemental Material, but, in this section, we use the more cumbersome tilde notation to indicate the different configurations of the sheet.}, which defines the stretches
\begin{align}
  &\tilde{f}_{s}(s) = \dfrac{\mathrm{d}\tilde{s}}{\mathrm{d}s}, &&\tilde{f}_{\phi}(s) = \dfrac{\tilde{s}(s)}{s}.  \label{eq:stretches}
\end{align}
As the sheet deforms, normals to $\mathcal{S}$ need not \emph{a priori} remain normal to $\tilde{\mathcal{S}}$. Hence a point in $\mathcal{V}$ at transverse position~$\zeta$ will end up, in $\tilde{\mathcal{V}}$, at transverse position $\tilde{\zeta}(s,\zeta)$, and displaced by a distance $\epsilon \tilde{\varsigma}(s,\zeta)$ parallel to $\tilde{\mathcal{S}}$. The position of a point in~$\tilde{\mathcal{V}}$ is therefore
\begin{align}
  \vec{\tilde{r}}(s,\phi, \zeta) = \vec{\tilde{\rho}}(s,\phi) + \epsilon \tilde{\zeta}(s,\zeta)\vec{e_z} + \epsilon \tilde{\varsigma}(s,\zeta)\vec{e_r}(\phi) .  
\end{align}
Now, using commata to denote partial differentiation, we can calculate
\begin{subequations}\label{eq:tbdef}
\begin{align}
&\dfrac{\partial \vec{\tilde{r}}}{\partial s} = \bigl(\tilde{f}_{s} + \epsilon \tilde{\varsigma}_{,s} \bigr)\vec{e_r} + \epsilon \tilde{\zeta}_{,s} \vec{e_z}, \\
&\dfrac{\partial \vec{\tilde{r}}}{\partial \phi} = \bigl( \tilde{r} + \epsilon \tilde{\varsigma} \bigr)\vec{e_\phi},\\
& \dfrac{\partial \vec{\tilde{r}}}{\partial \zeta} = \epsilon \bigl( \tilde{\zeta}_{,\zeta} \vec{e_z} + \tilde{\varsigma}_{,\zeta} \vec{e_r} \bigr).
\end{align}
\end{subequations}
\subsubsection*{\textbf{Plastic deformation: intermediate configuration}}
In the general continuum framework for plasticity~\cite{lubliner,bonet}, similarly to the theory of morphoelasticity~\cite{goriely}, the deformation gradient tensor $\smash{\tens{\tilde{F}}}$ decomposes multiplicatively into a plastic deformation gradient $\tens{F}^{\mathbfsf{p}}$ and an elastic deformation gradient $\tens{F}$, viz., $\smash{\tens{\tilde{F}}=\tens{F}\tens{F}^{\mathbfsf{p}}}$. The plastic deformation gradient is associated with the deformation from the undeformed configuration into an intermediate, virtual configuration $\mathcal{V}^0$, which we define by prescribing
\begin{align}
    \tens{F}^{\mathbfsf{p}}=
    \begin{pmatrix}
        1 & 0 &  0 \\
        0 & f(s)& 0 \\
        0& 0 & 1
    \end{pmatrix},\label{eq:Fp}
\end{align}
which introduces the intercalation stretch $f(s)$. As discussed in the main text, the preferred length of the ring at position $s$ in the undeformed configuration is thus $2\pi sf(s)$ in the intermediate configuration. We note that this intermediate configuration is not in general compatible, i.e., cannot be embedded into three-dimensional Euclidean space~\cite{goriely}. With this definition, the scale factors in the intermediate configuration are
\begin{align}
&\chi^0_s=1,&&\chi^0_\phi=s\,f(s),&&\chi^0_\zeta=\epsilon,
\end{align}
and so the volume element is, using Eq.~\eqref{eq:dvundef},
\begin{align}
\mathrm{d}V^0 =\chi^0_{s}\chi^0_{\phi}\chi^0_{\zeta}\,\mathrm{d}s\,\mathrm{d}\phi\,\mathrm{d}\zeta = \epsilon s\,f(s)\,\mathrm{d}s\,\mathrm{d}\phi\,\mathrm{d}\zeta=f(s)\,\mathrm{d}V. \label{eq:dvi}
\end{align}

\subsubsection*{\textbf{Calculation of the deformation gradient tensors}}
The deformation gradient tensor $\tilde{\tens{F}}$ mapping the undeformed configuration $\mathcal{V}$ of the sheet to its deformed configuration $\tilde{\mathcal{V}}$ is defined to be~\cite{goriely} 
\begin{subequations}
\begin{align}
    \tens{\tilde{F}} = \operatorname{Grad}{\vec{\tilde{r}}} = \dfrac{1}{\chi_s^2} \dfrac{\partial \tilde{\boldsymbol{r}}}{\partial s} \otimes \frac{\partial \boldsymbol{r}}{\partial s}+\dfrac{1}{\chi_\phi^2} \dfrac{\partial \tilde{\boldsymbol{r}}}{\partial \phi} \otimes \dfrac{\partial \boldsymbol{r}}{\partial \phi}+\dfrac{1}{\chi_\zeta^2} \dfrac{\partial \tilde{\boldsymbol{r}}}{\partial \zeta} \otimes \dfrac{\partial \boldsymbol{r}}{\partial \zeta}.
\end{align}
From Eqs.~\eqref{eq:drundef} and \eqref{eq:tbdef}, it follows that, with respect to $\tilde{\mathcal{B}} \otimes \mathcal{B}$, 
\begin{align}    
    \tens{\tilde{F}}=
    \begin{pmatrix}
    \tilde{f}_{s} + \epsilon \tilde{\varsigma}_{,s} & 0 & \tilde{\varsigma}_{,\zeta}\\
    0 & \tilde{f}_{\phi} & 0 \\
    \epsilon\tilde{\zeta}_{,s}& 0 & \tilde{\zeta}_{,\zeta}
    \end{pmatrix}. \label{eq:F}
    \quad
\end{align}
\end{subequations}
From this and Eq.~\eqref{eq:Fp}, we compute
\begin{align}
    \tens{F} = \tens{\tilde{F}}\left(\tens{F}^{\mathbfsf{p}}\right)^{-1} = 
    \begin{pmatrix}
         \tilde{f}_{s} + \epsilon\tilde{\varsigma}_{,s} & 0 & \tilde{\varsigma}_{,\zeta}\\
        0 & \dfrac{\tilde{f}_{\phi}}{f} & 0 \\
        \epsilon\tilde{\zeta}_{,s}& 0 & \tilde{\zeta}_{,\zeta}
    \end{pmatrix}. \label{eq:Fe}
    \quad
\end{align}

\subsection{Elasticity of the sheet}
We assume that the elastic sheet is made of an incompressible neo-Hookean hyperelastic material~\cite{goriely}, so that the elastic part of the energy is~\footnote{In this section of this Supplemental Material, we use $\mathcal{E}$ and $e$ to denote the elastic part of the energy and the corresponding three-dimensional energy density; in its other sections and in the main text, these symbols denote the total energy and the effective two-dimensional elastic energy density. We resort to this duplication to simplify our notation.}
\begin{subequations}
\begin{align}
\mathcal{E}=\iiint_{\mathcal{V}^0}{e\,\mathrm{d}V^0},\label{eq:energy}
\end{align}
in which the energy density is
\begin{align}
e = \dfrac{C}{2}(\mathcal{I}_1 - 3), \label{eq:e}
\end{align}
\end{subequations}
with $C>0$ a material parameter and $\mathcal{I}_1 = \operatorname{tr}{\tens{C}}$ the first invariant of the right Cauchy--Green tensor $\tens{C}=\tens{F}^\top\tens{F}$~\cite{goriely}. We stress that the integral is over the intermediate configuration $\mathcal{V}^0$ of the sheet, with volume element $\mathrm{d}V^0$ given by Eq.~\eqref{eq:dvi}. Assuming the simplest, neo-Hookean constitutive relation does not restrict our analysis: general hyperelastic constitutive relations are known to reduce to the neo-Hookean one for thin shells~\cite{erbay97,dervaux08,haas21}.

In the deformed configuration, the force on an area element $\mathrm{d}\tilde{S}$ with unit normal $\vec{\tilde{m}}$ is $ \tens{T} \vec{\tilde{m}}\,\mathrm{d}\tilde{S}$, where $\tens{T}$ is the Cauchy stress tensor~\cite{goriely,ogden}. For the neo-Hookean material assumed in Eq.~\eqref{eq:e},
\begin{align}
  \tens{T} = C\bigl(\tens{F}\tens{F}^\top - p\tens{I}\bigr),  \label{eq:Cauchytens}
\end{align}
in which $p$ is the Lagrange multiplier that imposes the incompressibility condition $\det{\tens{F}}=1$ and $\tens{I}$ is the identity~\cite{dervaux09}. The next step is to relate this force on an area element in the deformed configuration to the force on the corresponding area element $\mathrm{d}S$ in the undeformed configuration, which has unit normal $\vec{m}$. For this purpose, as in Ref.~\cite{haas21}, we introduce
\begin{align}
\tens{P} = \tilde{\mathcal{J}}\tens{T}\tens{\tilde{F}}^{-\top}=C\tens{Q},\quad\text{with }\tens{Q}=\tilde{\mathcal{J}}\left[\tens{F}\bigl(\tens{F}^{\mathbfsf{p}}\bigr)^{-\top}-p\tens{\tilde{F}}^{-\top}\right],\label{eq:Q}
\end{align}
where $\tilde{\mathcal{J}}=\det{\tens{\tilde{F}}}=\det{\tens{F}}\det{\tens{F}^{\mathbfsf{p}}}=f$.
Using Nanson's relation $\vec{\tilde{m}}\,\mathrm{d}\tilde{S} = \tilde{\mathcal{J}}\tens{\tilde{F}}^{-\top}\vec{m}\,\mathrm{d}S$~\cite{goriely,ogden}, we obtain $\tens{T}\vec{\tilde{m}}\,\mathrm{d}\tilde{S}=\tens{P}\vec{m}\,\mathrm{d}S$, and hence, similarly to the derivation of the Cauchy equation of classical elasticity~\cite{goriely,ogden} and to Ref.~\cite{haas21}, the equation
\begin{subequations}
\begin{equation}
\operatorname{Div}{\tens{Q}^{\top}} = \vec{0}, \label{eq:Cauchyeqn}
\end{equation}
determining the configuration of the sheet minimising the elastic energy in Eq.~\eqref{eq:energy}. Since $\mathcal{B}$ is independent of the transverse coordinate $\zeta$, this separates into
\begin{align}
\dfrac{(\tens{Q}\vec{e_z})_{,\zeta}}{\varepsilon}+\vec{\nabla}\cdot\tens{Q}^\top=\vec{0},\label{eq:Cauchy2}
\end{align}
\end{subequations}
in which the nabla operator denotes the gradient within the plane $\zeta=0$.

Equation~\eqref{eq:Cauchyeqn} is to be solved subject to force-free boundary conditions at the top and bottom surfaces of the sheet, which read $\tens{T}^\pm\vec{\tilde{n}^\pm}=\vec{0}$~\cite{goriely}, where $\vec{\tilde{n}^\pm}$ are the normals to these surfaces, and $\tens{T}^\pm$ are the Cauchy tensors evaluated there. By the above, these are equivalent to $\tens{Q}^\pm\vec{e_z}=\vec{0}$, where $\tens{Q}^\pm$ are evaluated on $\zeta=\pm h/2$.

\subsection{Asymptotic expansion}
We can now derive the effective elastic energy density of the sheet by asymptotic expansion for small deformations. We start by defining the elastic strains $E_s,E_\phi$ to match Eq.~(3) of the main text by writing
\begin{align}
   \tilde{f}_{s} = 1 + \epsilon E_{s}, && \tilde{f}_{\phi} = f  (1 + \epsilon E_{\phi}).
\end{align}
We introduce expansions
\begin{subequations}
\begin{align}
    \tilde{\zeta} = \tilde{\zeta}_{(0)} + \epsilon \tilde{\zeta}_{(1)} + O\bigl(\epsilon^{2}\bigr),  &&   \tilde{\varsigma}  = \tilde{\varsigma}_{(0)} + O(\epsilon) ,
\end{align}
and
\begin{align}
    \tens{Q}=  \tens{Q}_{\mathbfsf{(0)}} + \epsilon  \tens{Q}_{\mathbfsf{(1)}} + O\bigl(\epsilon^{2}\bigr),  &&  p  = p_{(0)} + O(\epsilon) .
 \end{align}
\end{subequations}
At leading order, Eq.~\eqref{eq:Cauchy2} is $(\tens{Q}_{\mathbfsf{(0)}}\vec{e_z})_{,\zeta}=\vec{0}$. This means that $\tens{Q}_{\mathbfsf{(0)}}\vec{e_z}$ is independent of $\zeta$, so the boundary conditions become $\vec{0}=\tens{Q}_{\mathbfsf{(0)}}\vec{e_z}+O(\varepsilon)$. Expanding definition~\eqref{eq:Q} using Eqs.~\eqref{eq:Fp}, \eqref{eq:F}, and \eqref{eq:Fe}, we find, at leading order, the differential equations
\begin{subequations}
\begin{align}
&\tilde{\varsigma}_{(0),\zeta}=0,&&\tilde{\zeta}_{(0),\zeta}-\frac{p}{\tilde{\zeta}_{(0),\zeta}}=0.
\end{align}
The expansion of the incompressibility condition $\det{\tens{F}}=1$ to leading order leads to
\begin{align}
\tilde{\zeta}_{(0),\zeta}=1.
\end{align}
\end{subequations}
By definition of the midsurface, $\tilde{\zeta}_{(0)}=\tilde{\varsigma}_{(0)}=0$ on $\zeta=0$, so these equations integrate to
\begin{align}
&\tilde{\zeta}_{(0)}=\zeta,&&\tilde{\varsigma}_{(0)}=0,&&p_{(0)}=1.
\end{align}
As noted in Ref.~\cite{haas21}, we need not continue the explicit expansion beyond this order. Rather, it is sufficient to consider a formal expansion
\begin{widetext}
\vspace{-15pt}
\begin{align}
    \tens{F} =
    \begin{pmatrix}
        1 + \epsilon a_{(1)} + \epsilon^{2} a_{(2)} + O\bigl(\epsilon^{3}\bigr)& 0 & \epsilon v_{(1)} + O\bigl(\epsilon^{2}\bigr) \\
        0 & 1 + \epsilon b_{(1)} + \epsilon^{2} b_{(2)} + O\bigl(\epsilon^{3}\bigr)  & 0 \\
        \epsilon w_{(1)} + O\bigl(\epsilon^{2}\bigr)& 0 & 1 + \epsilon c_{(1)} + \epsilon^{2} c_{(2)} + O\bigl(\epsilon^{3}\bigr)
    \end{pmatrix}, \label{eq:ExpansionF}
\end{align}
consistent with Eq.~\eqref{eq:Fe}. By direct computation, $a_{(1)}=E_s$, $b_{(1)}=E_\phi$. The incompressibility condition becomes
\begin{align}
      1=\det{\tens{F}} = 1 + \epsilon(a_{(1)}+ b_{(1)}+ c_{(1)} ) + 
      \epsilon^{2} (a_{(2)}+ b_{(2)}+ c_{(2)} + a_{(1)}b_{(1)} +  b_{(1)}c_{(1)} +  c_{(1)}a_{(1)} -  v_{(1)}w_{(1)}) + O\bigl(\epsilon^{3}\bigr).\label{eq:ExpansiondetF}
    \end{align}
\end{widetext}
From this, we obtain
\begin{subequations}\label{eq:c1c2}
\begin{align}
c_{(1)} &= - a_{(1)}- b_{(1)}, \\
c_{(2)} &=  a_{(1)}^{2}+ a_{(1)}b_{(1)}+ b_{(1)}^{2} - a_{(2)} - b_{(2)} + v_{(1)} w_{(1)}.   
\end{align}    
\end{subequations}
We can now compute, using Eq.~\eqref{eq:Fp},
\begin{align}
     \tens{\tilde{F}} = \tens{F} \tens{F}^{\mathbfsf{p}} =
     \begin{pmatrix}
         1 + O\bigl(\epsilon\bigr)   & 0 &  \epsilon v_{(1)} + O(\epsilon^{2} \bigr)\\
         0 &  f + O\bigl(\epsilon\bigr)  & 0 \\
         \epsilon  w_{(1)} + O\bigl(\epsilon^{2} \bigr)  & 0 &1 + O\bigl(\epsilon\bigr) 
     \end{pmatrix}.
\end{align}
Substituting $ p = 1 + O(\epsilon)$ in Eq.~\eqref{eq:Q} then leads to 
 \begin{align}
     \tens{Q} = \begin{pmatrix}
        O(\epsilon \bigr)  & 0 &  \epsilon f(v_{(1)} +  w_{(1)}) \\
        0& O(\epsilon \bigr) & 0 \\
        \epsilon f(v_{(1)} +  w_{(1)}) & 0 & O(\epsilon \bigr)
    \end{pmatrix}. 
 \end{align}
Since $\tens{Q}_{\mathbfsf{(0)}}=\mathbfsf{0}$, Eq.~\eqref{eq:Cauchy2} becomes $(\tens{Q}_{\mathbfsf{(1)}}\vec{e_z})_{,\zeta}=\vec{0}$, whence, similarly to before, $\tens{Q}_{\mathbfsf{(1)}}\vec{e_z} = \vec{0}$. This implies
\begin{align}
   w_{(1)} =-v_{(1)}.\label{eq:w1}
\end{align}
Using Eqs.~\eqref{eq:c1c2} and \eqref{eq:w1}, we can simplify the asymptotic expansion of the right Cauchy--Green tensor $\tens{C} = \tens{F}^\top\tens{F} $ and hence compute
\begin{align}
 \mathcal{I}_{1} = \operatorname{tr}{\tens{C}}= 3 + 4 \epsilon^{2}\bigl(a_{(1)}^{2}+ a_{(1)}b_{(1)}+ b_{(1)}^{2}\bigr) + O\bigl(\epsilon^{2} \bigr).\label{eq:I1}
\end{align}
To leading order and using Eq.~\eqref{eq:dvi}, the elastic energy in Eq.~\eqref{eq:energy} becomes
\begin{align}
\mathcal{E}=  \iint_{\mathcal{S}}{\hat{e}\,s\,\mathrm{d}s\,\mathrm{d}\phi} = 2\pi\int_{\mathcal{C}}{\hat{e}\,s\,\mathrm{d}s},
\end{align}
in which $\mathcal{C}$ is the line generating the undeformed midsurface~$\mathcal{S}$. The effective energy density is~\cite{Note2}
\begin{subequations}
\begin{align}
\hat{e}=\varepsilon\int_{-h/2}^{h/2}{e\,\mathrm{d}\zeta}.
\end{align}
On substituting Eq.~\eqref{eq:I1} into Eq.~\eqref{eq:e}, we find
\begin{align}
\hat{e}=\dfrac{4\varepsilon^3hC}{2}\bigl(E_s^2+E_sE_\phi+E_\phi^2\bigr)f.
\end{align}
\end{subequations}
This is the asymptotic form of Eq. (2) of the main text. The form of the equation in the main text is obtained by setting $\varepsilon=1$ or, more formally, by substituting $\varepsilon h\to h$, $\varepsilon E_s\to E_s$, $\varepsilon E_\phi\to E_\phi$, and identifying the effective elastic modulus $4hC\to C$.

\section{Derivation of the intercalation rates}
In this section, we derive the expressions for the intercalation rates used in the main text, and discuss their dependence on the asymptotic scaling of the effective temperature. We recall the model for the single-cell intercalation rates introduced in Eq. (6) of the main text,
\begin{align}
\rho_i^\pm\propto\exp{\left(-\dfrac{E_{\text{b},i}^\pm+\Delta E_i^\pm}{\theta}\right)},\label{eq:rho}
\end{align}
in which the effective temperature is $\theta=\vartheta\,\delta s^\tau$ and
\begin{align}
\Delta E_i^\pm=\delta s^3\left[\dfrac{\beta^2}{s}\dfrac{\partial^2e}{\partial f^2}\pm\beta\dfrac{\mathrm{d}}{\mathrm{d}s}\left(\dfrac{\partial e}{\partial f}\right)\right]_{\substack{s=s_i\\f=f_i}}+O\bigl(\delta s^4\bigr),\label{eq:DE}
\end{align}
as obtained in Eq. (7) of the main text. We will derive explicit expressions for $\partial e/\partial f$ and $\partial^2 e/\partial f^2$ below, after deriving the equations governing the elastic problem.

\subsection{High energy barrier: $\vec{E_\text{b}\gg\Delta E}$}
We begin with the case $E_\text{b}\gg\Delta E$ of a high energy barrier. As in the main text, we analyse the simplest model, of a constant energy barrier $E^\pm_{\smash{\text{b},i}}=B\,\delta s^2$, in which we have chosen $E_\text{b}=O\bigl(\delta s^2\bigr)$ to ensure $E_\text{b}\gg\Delta E=O\bigl(\delta s^3\bigr)$ from Eq.~\eqref{eq:DE}. (This model of a constant energy barrier is of course a simplification, since one expects the energy barrier to depend on details of the cell packing and cell mechanics.) As mentioned in the main text, different values of the exponent $\tau$ in the expression for the effective temperature $\theta=\vartheta\,\delta s^\tau$ yield different expressions for the effective intercalation rate $I_i\propto N_i\rho_i^+-N_{i+1}\rho_{i+1}^-$ at leading order, and hence different continuum models.

\subsubsection*{\textbf{``Cold'' intercalations:} $\vec{\tau>2}$}
If $\tau>2$, then, on expanding the expressions~\eqref{eq:rho} for the intercalation rates, we find 
\begin{align}
\rho_i^\pm\propto\exp{\left(-\dfrac{B}{\vartheta}\delta s^{-(\tau-2)}\right)}\to 0\quad\text{as }\delta s\to 0,
\end{align}
since $B>0$. Thus $I=0$, too. Physically, the energy barrier to intercalations cannot be overcome and hence there are no intercalations, so this case is uninteresting.

\subsubsection*{\textbf{``Hot'' intercalations:} $\vec{\tau<2}$}
If $\tau<2$, a similar expansion of the intercalation rates in Eq.~\eqref{eq:rho} gives
\begin{align}
\rho_i^\pm\propto 1-\dfrac{B}{\vartheta}\delta s^{2-\tau}+O\bigl(\delta s^{3-\tau},\delta s^{4-2\tau}\bigr).
\end{align}
In the continuum limit $\delta s\to 0$, $N_i\to n(s)=sf(s)$ from the definitions in the main text. Hence, 
on Taylor expanding $N_{i+1}=n(s_i)+n'(s_i)\delta s+O\bigl(\delta s^2\bigr)$, we infer
\begin{subequations}
\begin{align}
I_i\propto -n'(s_i)\delta s+O\bigl(\delta s^2,\delta s^{3-\tau},\delta s^{4-2\tau}\bigr),
\end{align}
with ${3-\tau},\;{4-2\tau}>1$ for $\tau<2$. Hence
\begin{align}
I\propto -\dfrac{\mathrm{d}n}{\mathrm{d}s}.
\end{align}
\end{subequations}
We note in passing that this illustrates that the proportionality constant must scale with $\delta s$ appropriately for the continuum limit to converge.

Thus, for $\tau<2$, intercalation is independent of the mechanics, and the system is driven entropically to the state in which $\mathrm{d}n/\mathrm{d}s=0$, i.e., each ring of cells surrounding the gap contains the same number of cells. This case is therefore not interesting, either.

\subsubsection*{\textbf{Intermediate case:} $\vec{\tau=2}$}
Any interesting behaviour must therefore arise in the intermediate case $\tau=2$, in which expansion of Eq.~\eqref{eq:rho} using Eq.~\eqref{eq:DE} yields
\begin{subequations}
\begin{align}
\rho_i^\pm &\propto \mathrm{e}^{-B/\vartheta}\left\{1-\dfrac{\delta s}{\vartheta}\left[\dfrac{\beta^2}{s}\dfrac{\partial^2e}{\partial f^2}\pm\beta\dfrac{\mathrm{d}}{\mathrm{d}s}\left(\dfrac{\partial e}{\partial f}\right)\right]\right\}+O\bigl(\delta s\bigr)^2.
\end{align}
Equivalently, this establishes an expansion
\begin{align}
\rho^\pm(s_i)  &= \rho_0 + \rho_1^\pm(s_i) \delta s + O\bigl(\delta s^2\bigr),
\end{align}
\end{subequations}
where, by comparison,
\begin{align}
&\rho_0=\mathrm{e}^{-B/\vartheta},&&\rho_1^+(s)-\rho_1^-(s)=-\dfrac{2\beta}{\vartheta}\dfrac{\mathrm{d}}{\mathrm{d}s}\left(\dfrac{\partial e}{\partial f}\right)\mathrm{e}^{-B/\vartheta}.\label{eq:rho0rho1}
\end{align}
Now, expanding the definition $I_i\propto N_i\rho_i^+-N_{i+1}\rho_{i+1}^-$,
\begin{subequations}
\begin{align}
I(s_i)&\propto n(s_i)\left[\rho_0+\rho_1^+(s_i)\delta s\right]\nonumber\\
&\qquad-\left[n(s_i)+n'(s_i)\delta s\right]\left[\rho_0+\rho_1^-(s_i)\delta s\right]+O\bigl(\delta s^2\bigr)\nonumber\\
&\propto\left\{n(s_i)\left[\rho_1^+(s_i)\!-\!\rho_1^-(s_i)\right]\!-\!n'(s_i)\rho_0(s_i)\right\}\delta s+O\bigl(\delta s^2\bigr).
\end{align}
Substituting from Eqs.~\eqref{eq:rho0rho1}, this yields
\begin{align}
I\propto -\left[\dfrac{\mathrm{d}n}{\mathrm{d}s}+\dfrac{2\beta}{\vartheta}n\dfrac{\mathrm{d}}{\mathrm{d}s}\left(\dfrac{\partial e}{\partial f}\right)\right]\mathrm{e}^{-B/\vartheta},\label{eq:HEB}
\end{align}
\end{subequations}
which is Eq.~(8) of the main text. The intercalation rate now depends on the mechanics, which leads to the interesting behaviour discussed in the main text. As noted there, the dependence of $I$ on $\mathrm{e}^{-B/\vartheta}$ merely rescales time, however.
\subsection{Fluidised tissue: $\vec{E_\text{b}\ll\Delta E}$}
We now turn to the fluidised case, in which we can neglect the energy barrier $E_\text{b}$ compared to $\Delta E$. Again, there are different cases depending on the exponent $\tau$. 
\subsubsection*{\textbf{``Hot'' intercalations: $\vec{\tau<3}$}}
In the case $\tau<3$, expanding the intercalation rates given by Eq.~\eqref{eq:rho} yields
\begin{align}
\rho_i^\pm\propto 1-\dfrac{\delta s^{3-\tau}}{\vartheta}(A_i\pm B_i)+O\bigl(\delta^{4-\tau},\delta s^{6-2\tau}\bigr),
\end{align}
where, from Eq.~\eqref{eq:DE},
\begin{align}
A_i=\left[\dfrac{\beta^2}{s}\dfrac{\partial^2e}{\partial f^2}\right]_{\substack{s=s_i\\f=f_i}},&&B_i=\left[\beta\dfrac{\mathrm{d}}{\mathrm{d}s}\left(\dfrac{\partial e}{\partial f}\right)\right]_{\substack{s=s_i\\f=f_i}}.\label{eq:AB}
\end{align}
Taylor expanding $N_{i+1}=n(s_i)+n'(s_i)\delta s+O\bigl(\delta s^2\bigr)$ again,
\begin{align}
I_i\propto -n'(s_i)\delta s-2B_in(s_i)\delta s^{3-\tau}+O\bigl(\delta s^2,\delta s^{4-\tau},\delta s^{6-2\tau}\bigr).
\end{align}
There are therefore three different subcases: First, if $\tau<2$, then this reduces to
\begin{subequations}
\begin{align}
I\propto -\dfrac{\mathrm{d}n}{\mathrm{d}s}\quad (\tau<2),
\end{align}
which is the uninteresting result that we also found in the high-energy barrier case for $\tau<2$: intercalation is independent of the mechanics and the system is driven entropically to the state in which all rings contain the same number of cells. Next, if $\tau=2$, then, substituting from Eq.~\eqref{eq:AB},
\begin{align}
I\propto -\left[\dfrac{\mathrm{d}n}{\mathrm{d}s}+\dfrac{2\beta}{\vartheta}n\dfrac{\mathrm{d}}{\mathrm{d}s}\left(\dfrac{\partial e}{\partial f}\right)\right]\quad (\tau=2),
\end{align}
which is, interestingly, Eq.~\eqref{eq:HEB} in the case of a vanishing energy barrier, $B=0$. The dynamics are therefore equivalent to the high-energy barrier dynamics, up to rescaling the intercalation rate or, equivalently, time. Finally, if $2<\tau<3$, then
\begin{align}
I\propto -n\dfrac{\mathrm{d}}{\mathrm{d}s}\left(\dfrac{\partial e}{\partial f}\right)\quad (2<\tau<3),
\end{align}
\end{subequations}
These are dynamics that do not arise in the high-energy barrier case. They are independent (up to rescaling the intercalation rate or time), however, of $\beta$ and $\vartheta$. We will not analyse these dynamics in detail.

\subsubsection*{\textbf{Intermediate case: $\vec{\tau=3}$}}
Instead, we will analyse the interesting dynamics of the intermediate case $\tau=3$ in detail. Now
\begin{align}
\rho_i^\pm\propto\exp{\left(-\left\{\dfrac{\beta}{\vartheta}\left[\dfrac{\beta}{s}\dfrac{\partial^2e}{\partial f^2}\pm\dfrac{\mathrm{d}}{\mathrm{d}s}\left(\dfrac{\partial e}{\partial f}\right)\right]\right\}\right)}+O(\delta s).
\end{align}
Hence $I_i\propto N_i(\rho_i^+-\rho_i^-)+O(\delta s)$, so
\begin{align}
I\propto-n\exp{\left(-\dfrac{\beta^2}{s\vartheta}\dfrac{\partial^2 e}{\partial f^2}\right)\sinh{\left(\dfrac{\beta}{\vartheta}\dfrac{\mathrm{d}}{\mathrm{d}s}\left(\dfrac{\partial e}{\partial f}\right)\right)}},
\end{align}
which is Eq.~(9) of the main text, where we analyse the dynamics of this intermediate case in detail.

\subsubsection*{\textbf{``Cold'' intercalations: $\vec{\tau>3}$}}
Finally, we discuss the case $\tau>3$. Since there is now no energy barrier to intercalations, intercalations are possible, but we expect only energetically favourable intercalations to happen. Indeed,
\begin{align}
\rho_i^\pm\propto \exp{\left(-\left\{\dfrac{\beta}{\vartheta}\left[\dfrac{\beta}{s}\dfrac{\partial^2e}{\partial f^2}\pm\dfrac{\mathrm{d}}{\mathrm{d}s}\left(\dfrac{\partial e}{\partial f}\right)\right]\right\}\delta s^{-(\tau-3)}\right)}.
\end{align}
Upon taking the continuum limit $\delta s\to 0$ and rescaling, we find
\begin{align}
\rho_i^\pm\propto\mathrm{H}\left(-\left[\dfrac{\beta}{s}\dfrac{\partial^2e}{\partial f^2}\pm\dfrac{\mathrm{d}}{\mathrm{d}s}\left(\dfrac{\partial e}{\partial f}\right)\right]\right),
\end{align}
where $\mathrm{H}$ is the Heaviside function. As noted in Eq.~(7) of the main text and as shown below, $\partial^2e/\partial f^2>0$, so at most one of $\rho_i^+$ and $\rho_i^-$ is nonzero. This means that, perhaps not unexpectedly, intercalations can be energetically favourable only in one direction. We thus obtain
\begin{align}
I\propto \left\{\begin{array}{cl}
n&\text{if }\dfrac{\mathrm{d}}{\mathrm{d}s}\left(\dfrac{\partial e}{\partial f}\right)<-\dfrac{\beta}{s}\dfrac{\partial^2e}{\partial f^2};\\
-n&\text{if }\dfrac{\mathrm{d}}{\mathrm{d}s}\left(\dfrac{\partial e}{\partial f}\right)>\dfrac{\beta}{s}\dfrac{\partial^2e}{\partial f^2};\\
0&\text{otherwise}.
\end{array}\right.
\end{align}
Again, these dynamics are not found in the high-energy barrier case. We note that they also introduce an explicit dependence on the aspect ratio $\beta$, but we will not analyse these dynamics in detail.

\section{Numerical methods}
In this final section, we discuss the numerical solution of the intercalation problem.
\subsection{Governing equations for the elastic problem}
We begin by deriving the boundary value problem that describes the deformed configuration $r(s)$ of the sheet. We vary the energy in Eq.~(1) of the main text, using the expression for the energy density in Eq.~(2) of the main text to find
\begin{align}
\dfrac{\delta \mathcal{E}}{2\pi C} &=  \int_{s_0}^{s_1}{\bigl[n_s(s)\,\delta E_s(s)+n_\phi(s)\,\delta E_\phi(s)\bigr]f(s)\,s\,\mathrm{d}s}\nonumber\\
&\qquad+ \dfrac{\Lambda}{C}\delta r(s_0),
\end{align}
where we have defined
\begin{align}
&n_s(s)=E_s(s)+\dfrac{E_\phi(s)}{2},&&n_\phi(s)=E_\phi(s)+\dfrac{E_s(s)}{2}.
\end{align}
The definitions in Eq.~(3) of the main text give ${\delta E_s(s)=\delta r'(s)}$, $\delta E_\phi(s)=\delta r(s)/[sf(s)]$. We now define the stresses
\begin{align}
&N_s=\dfrac{f(s)}{f_\phi(s)}n_s(s),&&N_\phi(s)=\dfrac{n_\phi(s)}{f_s(s)},
\end{align}
where we have introduced the stretches
\begin{align}
&f_s(s)=r'(s),&&f_\phi(s)=\dfrac{r(s)}{s},\label{eq:stretches2}
\end{align}
consistently with their definition in Eq.~\eqref{eq:stretches} during the derivation of the elastic energy density in the first section of this Supplemental Material. 
Hence 
\begin{align}
\dfrac{\delta\mathcal{E}}{2\pi C}&=\int_{s_0}^{s_1}{\bigl[N_s(s)r(s)\delta r'(s)+N_\phi(s)f_s(s)\delta r(s)\bigr]\,\mathrm{d}s}\nonumber\\
&\qquad+\dfrac{\Lambda}{C}\delta r(s_0)\nonumber\\
&=\int_{s_0}^{s_1}{\left[f_s(s)N_\phi(s)-\dfrac{\mathrm{d}}{\mathrm{d}s}\bigl(r N_s\bigr)\right]\,\mathrm{d}s}\nonumber\\
&\qquad+r(s_1)N_s(s_1)\delta r(s_1)+\left[\dfrac{\Lambda}{C}-r(s_0)N_s(s_0)\right]\delta r(s_0).
\end{align}
The deformed shape of the sheet is therefore described by
\begin{align}
    \dfrac{\mathrm{d} r}{\mathrm{d}s }  = f_s , &&
    \dfrac{\mathrm{d}N_s}{\mathrm{d}s } = \dfrac{f_s}{r} (N_\phi - N_s),\label{eq:DES}
\end{align} 
of which the first equation follows from definitions~\eqref{eq:stretches2} and the second from the variation. The variation also yields the boundary conditions: at the inner boundary, we impose
\begin{subequations}
\begin{align}
r(s_0)N_s(s_0)=\dfrac{\Lambda}{C}.
\end{align}
At the outer boundary, there are two natural choices,
\begin{align}
&r(s_1)=s_1&&\text{or}&&N_s(s_1)=0.\label{eq:bcchoices}
\end{align}
\end{subequations}
In the main text, we choose the first of these conditions; we discuss the second choice in \textwholefigref{figS2} and at the end of this section, where we will see that this only affects results at a quantitative, rather than qualitative level.

\subsubsection*{\textbf{Numerical solution of the elastic problem}}
As announced in the main text, at each time step, we solve these elastic governing equations, given the intercalation stretch $f(s)$ at this point in time, using the \texttt{bvp4c} function of \textsc{Matlab} (The MathWorks, Inc.).

During the numerical solution, we compute successively, from $r(s),N_s(s)$, and the fixed $f(s)$,
\begin{subequations}\label{eq:DErels}
\begin{align}
f_\phi(s)&=\dfrac{r(s)}{sf(s)},&E_\phi(s)&=f_\phi(s)-1,
\end{align}
followed by
\begin{align}
&E_s(s)=\dfrac{f_\phi(s)}{f(s)}N_s(s)-\dfrac{E_\phi(s)}{2},&&f_s(s)=E_s(s)+1,
\end{align}
and finally
\begin{align}
N_\phi(s)=\dfrac{1}{f_s(s)}\left[E_\phi(s)+\dfrac{E_s(s)}{2}\right].
\end{align}
\end{subequations}
In this way, we can compute the right-hand sides of Eqs.~\eqref{eq:DES}, so Eqs.~\eqref{eq:DErels} close the solution of the boundary-value problem.

\subsubsection*{\textbf{Calculation of derived quantities}}
We can also now derive the explicit expressions, given in Eq.~(7) of the main text, for the quantities derived from the energy density that appear in the expressions for the intercalation rates in Eqs.~(8) and (9) of the main text derived above. From Eq.~(3) of the main text and Eq.~\eqref{eq:stretches2},
\begin{align}
&\dfrac{\partial E_s}{\partial f}=0,&&\dfrac{\partial E_\phi}{\partial f}=-\dfrac{f_\phi}{f^2}=-\dfrac{E_\phi+1}{f}.
\end{align}
Equation~(2) of the main text then yields
\begin{subequations}
\begin{align}
\dfrac{\partial e}{\partial f}&=\dfrac{C}{2}\left[\bigl(E_s^2+E_sE_\phi+E_\phi^2\bigr)+f(E_s+2E_\phi)\dfrac{\partial E_\phi}{\partial f}\right]\nonumber\\
&=\dfrac{C}{2}\bigl(E_s^2-E_\phi^2-E_s-2E_\phi\bigr),\\
\dfrac{\partial^2 e}{\partial f^2}&=-C(E_\phi+1)\dfrac{\partial E_\phi}{\partial f}=\dfrac{Cf_\phi^2}{f^3}=\dfrac{Cr^2}{s^2f^3}>0,
\end{align}    
\end{subequations}
as stated in Eq.~(7) of the main text. To obtain expressions for $I$ and $\mathrm{d}I/\mathrm{d}s$, from Eqs.~(8) and (9) of the main text, we need to compute
\begin{subequations}
\begin{align}
\dfrac{\mathrm{d}}{\mathrm{d}s}\left(\dfrac{\partial e}{\partial f}\right)&=\dfrac{C}{2}\left[(2E_s-1)\dfrac{\mathrm{d}E_s}{\mathrm{d}s}-2(E_\phi+1)\dfrac{\mathrm{d}E_\phi}{\mathrm{d}s}\right],\\
\dfrac{\mathrm{d}^2}{\mathrm{d}s^2}\left(\dfrac{\partial e}{\partial f}\right)&=\dfrac{C}{2}\left[2\left(\dfrac{\mathrm{d}E_s}{\mathrm{d}s}\right)^2+(2E_s-1)\dfrac{\mathrm{d}^2E_s}{\mathrm{d}s^2}-2\left(\dfrac{\mathrm{d}E_\phi}{\mathrm{d}s}\right)^2\right.\nonumber\\
&\hspace{15mm}\left.-2(E_\phi+1)\dfrac{\mathrm{d}^2E_\phi}{\mathrm{d}s^2}\right],\\
\dfrac{\mathrm{d}}{\mathrm{d}s}\left(\dfrac{\partial^2 e}{\partial f^2}\right)&=\dfrac{Cf_\phi}{f^3}\left(2\dfrac{\mathrm{d}f_\phi}{\mathrm{d}s}-\dfrac{3f_\phi}{f}\dfrac{\mathrm{d}f}{\mathrm{d}s}\right).
\end{align}
\end{subequations}
We can use governing equations~\eqref{eq:DES}, relations~\eqref{eq:DErels}, and their derivatives repeatedly to obtain explicit expressions for the right-hand sides of these results that allow us to compute them without numerical differentiation of any of these elastic quantities.

\begin{figure*}[t]
    \centering\includegraphics[width=\linewidth]{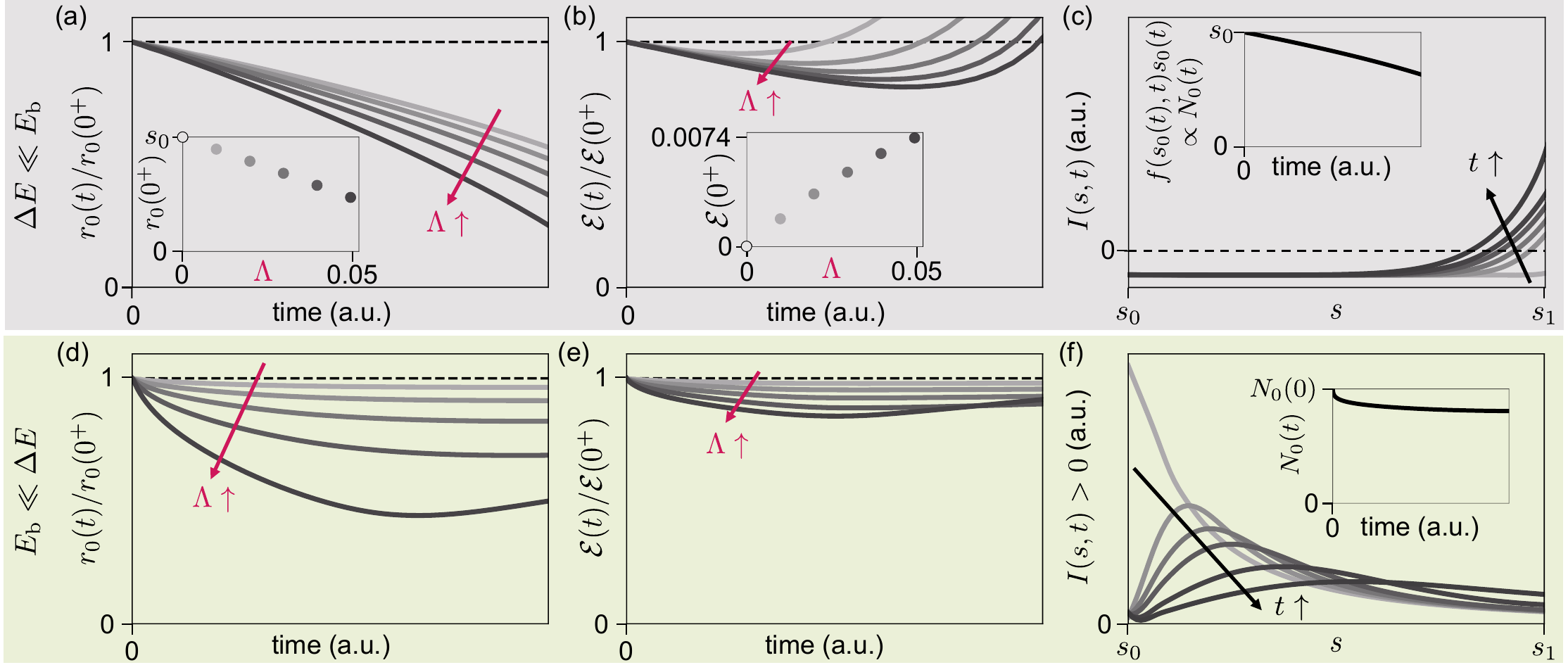}
    \caption{Dynamics of epithelial gap closure with a force-free outer boundary, similar to Fig.~4 of the main text. Top row: dynamics for a high energy barrier ($\Delta E\ll E_\text{b}$). (a) Plot of the inner radius $r_0(t)\equiv r(s_0(t),t)$, scaled with its initial contracted value $r_0(0^+)$ before intercalations, against time $t$, for increasing values of contractility $\Lambda$. Inset: plot of $r_0(0^+)$ against $\Lambda$. (b) Plot of energy $\mathcal{E}(t)$, scaled with $\mathcal{E}(0^+)$, against $t$, for increasing values of $\Lambda$. Inset: plot of $\mathcal{E}(0^+)$ against $\Lambda$. (c) Plot of the intercalation rate $I(s,t)$ against $s$, at increasing times $t$. Inset: decrease of $f(s_0(t),t)s_0(t)$, proportional to the number $N_0(t)$ of cells surrounding the gap, with time. Bottom row: analogous plots for the fluidised model ($\Delta E\gg E_{\text{b}}$). (d) Plot of $r_0(t)$. (e) Plot of $\mathcal{E}(t)$. (f) Plot of $I(s,t)$ against~$s$. Inset: plot of $N_0(t)$. Parameter values as in Fig.~4 of the main text: $s_0(0)=0.2$, $s_1(0)=1$, $C=\vartheta=1$, $B=0.1$, $\beta=0.02$, [panels (c), (f)] $\Lambda=0.02$. a.u.: arbitrary units.}
    \label{figS2}
\end{figure*}

\subsection{Numerical integration of the intercalation dynamics}
For numerical integration of the intercalation dynamics, given by Eqs. (4) of the main text, we define
\begin{subequations}
\begin{align}
s(\sigma,\tau)&=s_0(\tau)(1-\sigma)+s_1(\tau)\sigma,\\
t(\sigma,\tau)&=\tau.
\end{align}
Since
\begin{align}
\sigma(s,t)&=\dfrac{s-s_0(t)}{s_1(t)-s_0(t)},&\tau(s,t)&=t,
\end{align}
\end{subequations}
the inverse of this change of variables is well-defined. This allows us move the problem from $s\in[s_0(t),s_1(t)]$ to the fixed interval $\sigma\in[0,1]$.
\begin{widetext}
\noindent The chain rule yields
\begin{align}
\left(\dfrac{\partial f}{\partial t}\right)_s&=\left(\dfrac{\partial f}{\partial\tau}\right)_\sigma-\dfrac{\sigma\dot{s}_1(\tau)+(1-\sigma)\dot{s}_0(\tau)}{s_1(\tau)-s_0(\tau)}\left(\dfrac{\partial f}{\partial\sigma}\right)_\tau,&\left(\dfrac{\partial I}{\partial s}\right)_t&=\dfrac{1}{s_1(\tau)-s_0(\tau)}\left(\dfrac{\partial I}{\partial \sigma}\right)_\tau,\label{eq:cv}
\end{align}    
and so Eq.~(4a) of the main text becomes
\begin{align}
\left(\dfrac{\partial f}{\partial\tau}\right)_\sigma&=\dfrac{1}{s_1(\tau)-s_0(\tau)}\left\{-\dfrac{1}{s_0(\tau)(1-\sigma)+s_1(\tau)\sigma}\left(\dfrac{\partial I}{\partial \sigma}\right)_\tau+\left[\sigma\dot{s}_1(\tau)+(1-\sigma)\dot{s}_0(\tau)\right]\left(\dfrac{\partial f}{\partial\sigma}\right)_\tau\right\}.\label{eq:dfdtau}
\end{align}
\end{widetext}
As discussed in the main text, the dynamics of $\dot{s}_0(\tau),\dot{s}_1(\tau)$ are determined by
\begin{subequations}\label{eq:dsdtau}
\begin{align}
\dfrac{\mathrm{d}s_0}{\mathrm{d}\tau}=\left\{\begin{array}{cl}
\dfrac{I(0,\tau)}{s_0(\tau)f(0,\tau)}&\text{if }I(0,\tau)<0,\\
0&\text{otherwise},
\end{array}\right.\\
\dfrac{\mathrm{d}s_1}{\mathrm{d}\tau}=\left\{\begin{array}{cl}
\dfrac{I(1,\tau)}{s_1(\tau)f(1,\tau)}&\text{if }I(1,\tau)>0,\\
0&\text{otherwise}.
\end{array}\right.
\end{align}
\end{subequations}
We solve Eqs.~\eqref{eq:dfdtau} and~\eqref{eq:dsdtau} on a uniform grid $\{\sigma_i\}$ of $N_\text{mesh}$ points with spacing $\Delta\sigma$. We choose $N_\text{mesh}=20$ (in Fig.~4 of the main text and \textwholefigref{figS2}) or $N_\text{mesh}=30$ (in Fig.~5 of the main text). We perform the time integration of Eqs.~\eqref{eq:dfdtau} and~\eqref{eq:dsdtau} in \textsc{Matlab} using a custom fourth-order Runge--Kutta solver~\cite{abramowitz} with timestep $\Delta \tau$.

In the numerical solution, we use the values $\{f(\sigma_i)\}$ on the grid to define a spline that allows efficient evaluation of~$f$ by the \texttt{bvp4c} solver in the solution of the elastic problem. We differentiate this spline to evaluate the derivatives $\{f_{,\sigma}(\sigma_i)\},\{f_{,\sigma\sigma}(\sigma_i)\}$ on the mesh, which allows us to evaluate $I$ and $\partial I/\partial\sigma$ on the mesh without further numerical differentiation, as noted above.

To choose the timestep $\Delta\tau$, we impose a Courant--Friedrichs--Lewy condition~\cite{nr}: we compute an effective propagation speed $\{u_i\}=\{f_{,\tau}(\sigma_i)/f_{,\sigma}(\sigma_i)\}$ and choose $\Delta\tau$ to satisfy $\Delta\tau < C\,\Delta \sigma/\max{\{u_i\}}$, where $C=0.5$.

\subsubsection*{\textbf{Cell number conservation}}
If $\dot{s}_0(\tau)=0$ or $\dot{s}_1(\tau)=0$, then, as discussed in the main text, Eq.~(4a) changes to Eq.~(4b), which includes Dirac deltas at the boundaries to satisfy the condition of cell number conservation expressed by Eq.~(5a) of the main text near the inner boundary $s=s_0(t)$. 

In the numerical solution, we do not solve this equation including Dirac deltas, but rather impose cell number conservation on the mesh. Differentiating Eq.~(5a) of the main text, we obtain
\begin{align}
\int_{s_0(t)}^S{\dfrac{\partial f}{\partial t}s\,\mathrm{d}s}-\dot{s}_0(t)f\bigl(s_0(t),t\bigr)s_0(t)+\beta I(S)=0.\label{eq:dmass}
\end{align}
If $\dot{s}_0(t)=0$, then we split the integral at the second mesh point $s=s_0+\Delta s$ (i.e., $\sigma=\Delta\sigma$), and evaluate the first part using the trapezoidal rule and the second part using Eq.~(4a) to find
\begin{align}
\int_{s_0}^S{\dfrac{\partial f}{\partial t}s\,\mathrm{d}s}&=\int_{s_0}^{s_0+\Delta s}{\dfrac{\partial f}{\partial t}s\,\mathrm{d}s}+\int_{s_0+\Delta s}^S{\dfrac{\partial f}{\partial t}s\,\mathrm{d}s}\nonumber\\
&=\dfrac{\Delta s}{2}\left[s_0\dfrac{\partial f}{\partial t}(s_0)+(s_0+\Delta s)\dfrac{\partial f}{\partial t}(s_0+\Delta s)\right]\nonumber\\
&\qquad+\int_{s_0+\Delta s}^S{\left(-\beta\dfrac{\partial I}{\partial s}\right)\,\mathrm{d}s}\nonumber\\
&=\dfrac{\Delta s}{2}\left[s_0\dfrac{\partial f}{\partial t}(s_0)-\beta\dfrac{\partial I}{\partial s}(s_0+\Delta s)\right]\nonumber\\
&\qquad-\beta I(S)+\beta I(s_0+\Delta s),
\end{align}
having using Eq.~(4a) again to obtain the final equality. On substituting this result into Eq.~\eqref{eq:dmass}, we obtain
\begin{subequations}
\begin{align}
\dfrac{\partial f}{\partial t}(s_0)=\dfrac{\beta}{s_0}\left[\dfrac{\partial I}{\partial s}(s_0+\Delta s)-\dfrac{2I(s_0+\Delta s)}{\Delta s}\right],
\end{align}
which thus replaces Eq.~(4a) when $\dot{s}_0(t)=0$. Similarly, for ${\dot{s}_1(t)=0}$, we find
\begin{align}
\dfrac{\partial f}{\partial t}(s_1)=\dfrac{\beta}{s_1}\left[\dfrac{\partial I}{\partial s}(s_1-\Delta s)+\dfrac{2I(s_1-\Delta s)}{\Delta s}\right].
\end{align}
\end{subequations}
We translate these equations to the variables $(\sigma,\tau)$ used in the numerical solution using Eqs.~\eqref{eq:cv}.
\subsection{Dynamics of epiboly: Effect of boundary conditions}
Figure~4 of the main text shows the dynamics for a fixed outer boundary, $r(s_1(t),t)=s_1(0)$. From Eq.~\eqref{eq:bcchoices}, the alternative is the force-free boundary condition $N_s(s_1(t),t)=0$. Results for this boundary conditions are shown in \textwholefigref{figS2} and only differ at a quantitative level: the boundary conditions thus have but a slight effect on the dynamics of epiboly.

\bibliography{references}